\documentclass[aps,showpacs,showkeys,prc,groupedaddress,twocolumn,superscriptaddress]{revtex4}

\usepackage{graphicx}

\begin{document}

\preprint{BB0704}

\title{MACE --- Mach cones in heavy ion collisions}

\author{Bj\o{}rn B\"auchle}
\email{baeuchle@th.physik.uni-frankfurt.de}
\affiliation{Institut f\"ur Theoretische Physik, Johann Wolfgang
Goethe-Universit\"at, Max-von-Laue-Str.~1, D-60438 Frankfurt am Main,
Germany}

\author{Laszlo Csernai}
\affiliation{Section for Theoretical Physics, Departement of Physics,
University of Bergen, All\'egaten 55, 5007 Bergen, Norway}
\affiliation{KFKI Research Institute for Particle and Nuclear Physics, P.O.
Box 49, 1525 Budapest, Hungary}

\author{Horst St\"ocker}
\affiliation{Institut f\"ur Theoretische Physik, Johann Wolfgang
Goethe-Universit\"at, Max-von-Laue-Str.~1, D-60438 Frankfurt am Main,
Germany}
\affiliation{Frankfurt Institute for Advanced Studies, Johann Wolfgang
Goethe-Universit\"at, Max-von-Laue-Str.~1, D-60438 Frankfurt am Main,
Germany}
\affiliation{Gesellschaft f\"ur Schwerionenforschung (GSI), D-64220
Darmstadt, Germany}

\date{\today}

\begin{abstract}

We study the propagation of sound-like perturbations created by a jet moving
with supersonic velocity through the quark-gluon-plasma created in heavy-ion
reactions within the model MACE (MAch Cone Evolution). Predictions for
heavy-ion reactions at RHIC energies (Au+Au-collisions) and for Pb+Pb
reactions at the LHC ($\sqrt s=5.5A$~TeV) are presented and potential
observations by the STAR, PHENIX and ALICE experiments are discussed. 

\end{abstract}


\maketitle

\section{Introduction}

First experimental data at the Relativistic Heavy Ion Collider (RHIC) have
shown the disappearance of backward correlations in hard processes at
central Au+Au-Collisions for
high-$p_\bot$-particles~\cite{Adler:2002tq,Adams:2003im}. Correlation data
with lower $p_\bot$-threshold for associated particles show a broader peak
or even a double-peaked structure in the backward region
\cite{Wang:2004kf,Adams:2005ph,Jacak:2005af,Wang:2005hp,Ajitanand:2005xa,Wang:2006xq,Molnar:2007wy}.
Three-particle-correlations~\cite{Ulery:2006ix} seem to support the theory
of conical emission~\cite{Ajitanand:2005xa,Wang:2006ig,Ulery:2007zb}.
Recently, the observations have been confirmed by preliminary data from the
CERES-experiment at the Super Proton Synchrotron (SPS) \cite{Kniege:2007mp}.

Conical structures in heavy ion collisions have been postulated first by
Scheid and Greiner already in 1974~\cite{Scheid:1974yi} and later also by
Hofmann, Baumgardt, St\"ocker and collaborators in the second half of the
1970s~\cite{Hofmann:1975by,Hofmann:1975cn,Baumgardt:1975qv,Hofmann:1976dy,Hofmann:1976gt,Hofmann:1976qd,Stoecker:1977ki,Stoecker:1979iv,Stoecker:1979wq,Stoecker:1980pi,Stoecker:1980vf}.
In that time, strong shock waves in cold nuclear
matter were discussed. New interest has been rising in this field in recent years,
triggered by the above-named experimental results and new theoretical
work from St\"ocker and Casalderrey-Solana \textit{et al.}\ in
2004~\cite{Stoecker:2004qu,CasalderreySolana:2004qm}, and lots of further
considerations in the years
since~\cite{Chaudhuri:2005vc,Satarov:2005mv,Antinori:2005tu,Renk:2005ta,Renk:2005si,Renk:2005rq,CasalderreySolana:2005rf,Renk:2006mv,Renk:2006es,Renk:2006pw,Shuryak:2006ii,Chaudhuri:2006qk,Shuryak:2007hd,Renk:2007qe,CasalderreySolana:2007km,Renk:2007rv,Stoecker:2007su,Chaudhuri:2007gq,Chaudhuri:2007vc}.

The common idea is that an ultra-relativistic jet
which is moving through an extended Quark-Gluon-Plasma will cause some kind
of perturbation, maybe sound-like, and thereby lose its energy. Then, it
cannot be seen in the experiments anymore. Since the speed of sound is
considerably smaller than the speed of light, the perturbations, if indeed
sound-like, should interfere constructively along a cone, the so-called Mach
cone. 

The term ``backward hemisphere'' is meant relative to one high-energetic
trigger-jet. The jet that creates the conical structure is supposed to be
opposite to the trigger, i.e.\ in exact backward direction.

The opening angle $\beta$ of this cone is in static medium given by
\begin{equation}\label{eq:sinangle} \beta = \sin^{-1}
\frac{c_\mathrm{S}}{v}\quad, \end{equation}
where $c_\mathrm{S}$ is the speed of sound in the medium and $v$ is the
velocity of the jet. The cone is supposed to directly emit (or indirectly enhance emission) of
particles. Those particles will be found in a circle around the direction of
the jet at
\begin{equation}\label{eq:cosangle} \alpha = \cos^{-1}
\frac{c_\mathrm{S}}{v}\quad.  \end{equation}

In two-particle correlations this scenario will manifest itself as a
double-peaked structure in the backward-hemisphere.

The case of non-static medium has been studied in several publications as
well for homogeneous, non-static medium as for simplified of full
3+1-dimensional hydrodynamical evolutions of the underlying medium. All of
the latter works have described the sound waves and Mach cones within the
framework of hydrodynamics and, for that reason, together with the
underlying medium. 

In the presented work, we develop a model that propagates sound waves
through the velocity field as created by applying the hydrodynamical
equations and discovers Mach cone structures automatically. Backreaction of
jet and sound waves onto the medium is neglected.

\section{Hydrodynamics}

A model widely used to describe heavy-ion reactions is fluid- or
hydrodynamics. It has been predicted as a key mechanism for the creation of
hot and dense matter very early \cite{Hofmann:1975by,Hofmann:1976dy}. Using
this approach one assumes that the matter described is, at least nearly, in local thermal
equilibrium.
This is only justified after a short period of time. After that, a locally
equilibrated system may be formed. The part of the reaction before
equilibration cannot be described by usual (1-fluid-) hydrodynamics.

Therefore, an additional model has to be applied for the creation of the
first equilibrated state, the so-called initial state. This can in principle
be any kind of non-equilibrium model. The creation of the first equilibrated
state can take different times, depending on the mechanism with which it is
reached. Times of the order of $\tau_0 \approx 1$~fm are reasonable.

When the initial state is defined, hydrodynamics start to work. Depending on
number and kind of the assumed symmetries in the initial state the fluid
development may be solvable analytically or only numerically. Within this
stage of the calculations, one has to assume an equation of state (EoS). It
must also be chosen if the evolution describes a perfect fluid that is
perfectly equilibrated at any point or if small deviations are endorsed. In
the latter case, several additional parameters like the heat conductivity,
shear- and bulk-viscosity have to be introduced.

In the case of a perfect liquid, there are six equations which have to be
solved. Among them are one equation for the conservation of the baryon
number density and four for the conservation of energy and momentum:
\begin{eqnarray}\label{eq:hydro_contin1} \partial_\mu N^\mu =&
0\\\label{eq:hydro_contin2} \partial_\mu T^{\mu\nu} =& 0 & , \end{eqnarray}
where the interesting variables are given by
\begin{eqnarray}\label{eq:N_def} N^\mu = &\varrho u^\mu \\\label{eq:T_def}
T^{\mu\nu} = &(\varepsilon + P) u^\mu u^\nu - P g^{\mu\nu} & .
\end{eqnarray}
with the local rest frame baryon density $\varrho$, the local rest frame
energy density $\varepsilon$, the pressure $P$ and the flow velocity
$u^\mu$.

The sixth equation needed is the EoS which gives a connection between
pressure and the densities:
\begin{equation}\label{eq:eos_general} P = P \left ( \varepsilon,\,\varrho
\right ) \quad.  \end{equation}

Within the hydrodynamical framework the propagation of sound waves can be
described. The speed of sound $c_\textrm{S}$ can here be derived from the
EoS as
\begin{equation}\label{cs} c_\textrm{S}^2 = \frac{\partial P}{\partial
\varepsilon} \quad.\end{equation}

\section{Ingredients}

The model developed for this work is independent of the model used
for the hydrodynamical evolution. This is because the hydro-evolution is
only used as an input for the model, not an inherent part of it. Indeed, any model that
produces a velocity field can be used. However, of course, a different input
will alter the results.

Here, we use the Particle in Cell Code (PiC) which bases on the method with
the same name developed by Amsden and Harlow in the early 1960s
\cite{Amsden:1968zz,Amsden:1978zz} and upgraded to ultra-relativistic
energies by Nix and Strottman in the 80s and 90s
\cite{Clare:1986qj,Strottman:1989zz,Amelin:1991kb}. The method takes
advantage of a combination of Eulerian and Lagrangian solution methods; the
pressure is calculated on an Eulerian grid (where the cells are fixed in
coordinate space), while all current transfers are
calculated in a larger number of Lagrangian cells (where the cells comoving
with the fluid).

The initial state for the hydro-evolution is calculated in an effective 1D
Yang-Mills coherent field model, where the transverse plane is split up to
``streaks'' like in the ``Firestreak''-model of the 1970s
\cite{Magas:2000jx,Magas:2002ge}. During the evolution, an ideal gas EoS is
used, so that the pressure $P$ is connected to the energy density
$\varepsilon$ in a very simple way:
\begin{equation}\label{eq:eos} P = 1 / 3 \,\varepsilon\quad, \end{equation}
and the speed of sound is given as
\begin{equation}\label{eq:cs} c_\textrm{S} = 1 / \sqrt{3} \approx 0.577
\quad.  \end{equation}

This corresponds to a first-order approximation onto a Quark-Gluon-Plasma
(QGP). The fact that there is no phase transition is justified because the
main part of sound wave evolution will take place in this first, hot and
dense stage of the reaction.

\section{MACE}

The model ``MAch Cones Evolution'' (MACE) is built to model the propagation
of sound waves through a realistic, relativistic medium. Our approach is to
take an existing hydro-evolution and impose a jet and the waves it creates
as perturbations. Therefore, no backreaction of jet or sound
waves onto the medium can be modelled.

\subsection{Jet}

Only the away-side-jet (the jet in the backward hemisphere) is considered, since the near-side-jet is not
affected by and does not affect the medium.

The jet starts at the first timestep calculated by the hydrodynamical code.
This is, depending on the system considered, after few $fm$ after the first
collisions.

The jet is created at a given point with a probability that is proportional
to the energy-density at that point. The jet's direction is totally random.
The only cut that is applied is the
exclusion of all jets that will not end up in the detector. That is, for
RHIC-data we take all jets with $|\eta| < 0.7$ (corresponding to the
STAR-TPC), and for LHC-data we cut at $|\eta| = 0.9$, which is the
acceptance of the ALICE-TPC.

The jet is considered to be high energetic enough not to change its velocity
and direction. Calculation of jet quenching is not within the scope of the
model. It hence propagates in a straight line with speed of light through
the medium. At every timestep, it excites sound waves at its position. These
waves are not enhanced in direction of the jet, but are undirected. The
direction of the jet will be used for correlation considerations in the end.
When out of the medium, the jet does not excite sound waves anymore.

\subsection{Sound waves}

No premature assumptions on the shape of a resulting mach front can be made
in an unforeseeable, inhomogeneous and non-statical medium. We will
therefore propagate the single elementary waves and identify the wave fronts
independent of the propagation. This allows for parts of the elementary
waves to become a part of a wave front only after some time and possible
deflection. Also, some parts may be swamped away out of the wave front. 

So, in order to model the sound waves we create a lot of logical particles,
so-called wave markers, at the position of the jet. The word ``logical''
refers to the fact that we do not assign any physical quantities to these
markers yet; they only represent the position and direction of the wave.
Random directions will be assigned to the markers, so that after a short
propagation in homogeneous medium one elementary wave should indeed be a
spherical wave. Such an elementary wave is created at each timestep at the
position of the jet. Between the timesteps all ``waves'', i.e.\ all wave
markers, are propagated. 

The propagation of the wave markers is straightforward: The only assumption
made is that they move with the speed of sound relative to the fluid
wherever they are. Their direction is adjusted by relativistically adding
their initial velocity, $\vec v$, to the flow-velocity of the underlying
medium at the current point, $\vec u$. This initial velocity $\vec v$ is
calculated by multiplying the direction of the wave marker $\hat v$ with the
speed of sound: $\vec v = c_{\rm S} \hat v$.

When a marker crosses the border of a fluid cell, then its current
propagation $\vec r\prime = \vec r \,+\, \vec v \cdot {\rm d}t$ will be
finished with the old velocity, i.e.\ no deflection happens at the border of
two cells. This is justified because the timestep ${\rm d}t$ is much smaller
than the dimensions of the cells ${\rm d}x$, ${\rm d}y$ and ${\rm d}z$.

\begin{figure} \includegraphics[width=.47\textwidth]{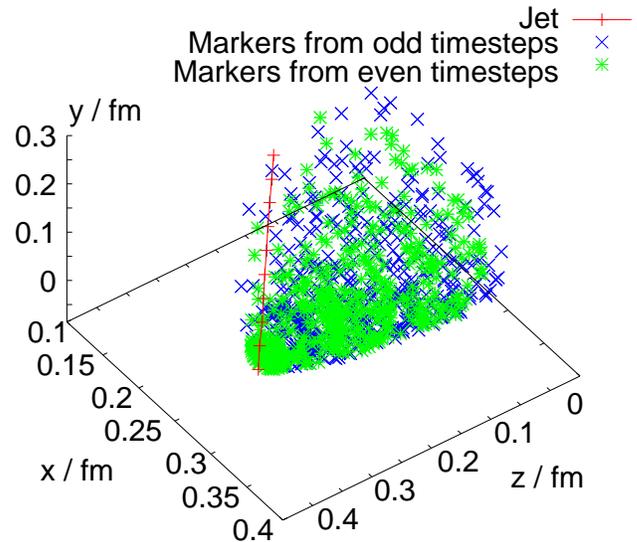}
\caption{(Color Online) Wave markers for a jet propagating through a
peripheral RHIC-collision after 10 timesteps. The jet was started near the
center of the collision with a rapidity of $y = 1.1$. The red line shows the
trajectory of the jet, the blue crosses and green stars show the wave
markers started at odd and even timesteps,
respectively.}\label{fig:propagate} \end{figure}

If a marker leaves the system, it is deleted. No particle emission at this
point is assumed, and the wave is also not reflected. Fig.\
\ref{fig:propagate} shows the position of markers after 10 timesteps of an
exemplified propagation (here, the data for a peripheral collision at RHIC have
been applied).

It should be noted that, by conserving the number of wave markers per
elementary wave, their ``amplitude'' will decrease with the
$1/r^2$-behaviour expected for spherical waves.

Since the underlying hydro-evolution is calculated using an EoS
with a constant speed of sound (see Eq.~(\ref{eq:cs})), it would be
inconsistent to use a density-dependent speed of sound for the propagation
of the sound waves. Therefore, propagation always happens with $c_{\rm S} =
1 / \sqrt{3}$.

\subsection{Mach cones}

In order to see the collective phenomenon of a Mach cone, the region
occupied by wave markers is reshaped with a set of lines, so-called wave
lines. They go along the surface of that region, starting at the position of
the jet. A handy analogy to this are longitudes going from the pole of the
earth along its surface and thereby reshaping it.

The wave lines are equally distributed azimuthally around the jet's
direction. Their nodes are positions of marker particles. By restricting the
possible nodes to those marker particles that are in the correct azimuthal
slice the azimuthal position of the wave lines is fixed.

The sheer surface of the area might be very distorted or irregular due to
numerical reasons. For instance, due to the finite density of timesteps, the
surface will contain large portions of the same elementary waves (in the
limit of infinitely many elementary waves each of these only contribute one
point on a circle in a plane orthogonal to the jet's direction).
Also, the random direction of the wave markers and their finite number may
cause distortion. It might be the case that in the area where a given
elementary wave should represent the surface there is no wave marker to
determine a possible node.

In MACE, the surface is smoothed to eliminate these effects. Care is taken
not to conceal physical effects during this procedure.

\begin{figure} \includegraphics[width=.47\textwidth]{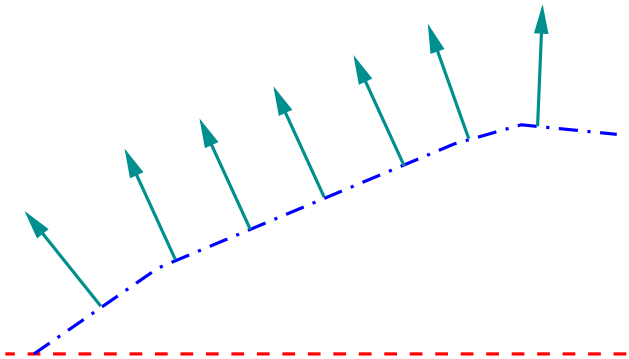}
\caption{(Color Online) Sketch on how the signals are created from the wave
lines. The red dotted line shows the jet trajectory, the blue dash-dotted
line a wave line.  The arrows are perpendicular to the wave line in constant
intervals along it.}\label{fig:signal}\end{figure}

When all nodes are collected and the lines are complete, particle flow is
considered to be orthogonal to the wave lines. The lines are now taken
\emph{as is}. To find the places on which to insert a signal we go along the
wave lines, starting from the jet, and insert a signal after fixed
distances. Momentum is inserted in the plane where the wave line is,
orthogonal to the jet direction (see Fig.~\ref{fig:signal}).

The signals are taken directly and mapped into correlation functions. There
is no new velocity or momentum profile created using signal and underlying
hydrodynamical fields. Therefore, background-subtracted
data are obtained more easlily. The correlation plots shown in section~\ref{sec:results} are
normalized to the number of signals. 

Obviously, the position and density of wave markers is totally lost during
this procedure. Instead, the density of wave lines is important. It is easy
to see that this drops with the inverse of the distance to the jet axis.
Since the amplitude of a Mach cone decreases with the same law, no
additional, artificial decreasing of the signal has to be performed.
Therefore, it is justified to keep the distance between to signals along the
wave lines constant.

The region right behind the jet is treated as any other region in the model.
Hence, the cones really have a sharp top. Non-linear behaviour and a rounder
shape in this region is not included.

\section{Results}\label{sec:results}

\begin{figure} \includegraphics[width=.47\textwidth]{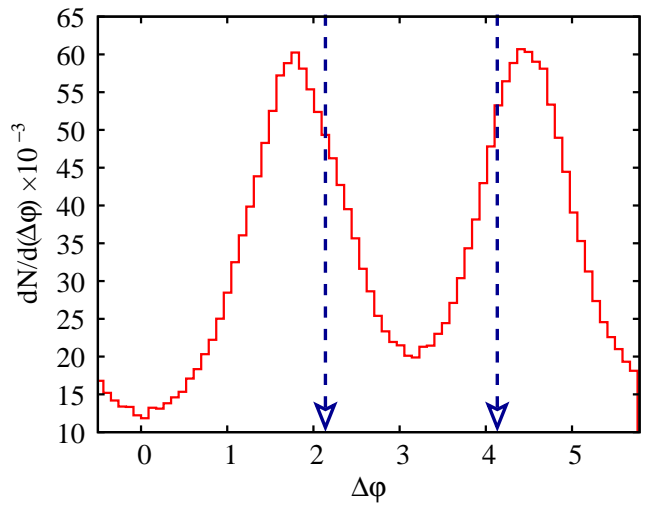}
\caption{(Color Online) Two-particle correlations (backward hemisphere) for
central gold-on-gold RHIC-collisions ($\sqrt{s_\textrm{NN}} = 130$~GeV and
$b = 0$).  The plot shows the average over 1\ 000 arbitrary jets. The arrows
show the ``perfect mach positions'' $\Delta \varphi = \pi \pm \cos^{-1}
c_\textrm{S}$.}\label{fig:rhic_00} \end{figure}

\begin{figure}
\includegraphics[width=.47\textwidth]{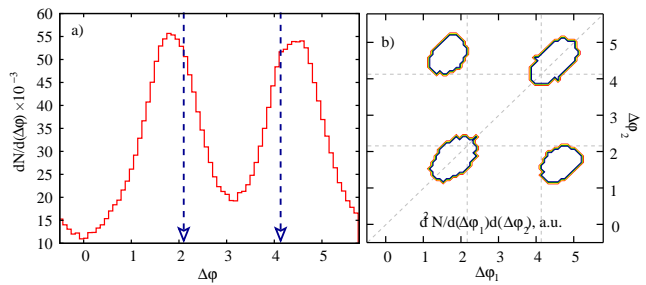} \caption{(Color Online) Two-
(a) and Three- (b) particle correlations (backward hemisphere) for central
lead-on-lead LHC-collisions ($\sqrt{s_\textrm{NN}} = 5.5$~TeV and $b = 0$).
Both plots show the average over 1\ 000 arbitrary jets. The arrows show the
``perfect mach positions'' $\Delta \varphi = \pi \pm \cos^{-1}
c_\textrm{S}$.}\label{fig:lhc_mb} \end{figure}

The correlation functions obtained with different velocity fields created
from different initial states show a very common shape. Both in central
RHIC- (see Fig.~\ref{fig:rhic_00}) and central LHC-Collisions (see
Fig.~\ref{fig:lhc_mb}) ($\sqrt{s_{\rm NN}} = 130$~GeV and $\sqrt{s_{\rm NN}}
= 5.5$~TeV, respectively) the averaged correlation functions show a clear
double-peaked structure in the backward hemisphere.  Also, in all cases
those two peaks are not at the expected positions $\Delta \varphi = \pi \pm
\cos^{-1} c_{\rm S}$, which would be $\Delta \varphi_1 = 2.2$
and $\Delta \varphi_2 = 4.1$. Instead, they are systematically at positions
further apart, so the angle to the backward direction is systematically
bigger. The magnitude of this shift is also very constant among the
considered cases; it is about $\delta \Delta \varphi \approx 0.2$.
Correlations for mid-central RHIC collisions (not shown) do not show a
significantly different picture.

Three-particle correlations (see Fig.~\ref{fig:lhc_mb}~(b)) show the
expected four-peaked structure, but also here it can be seen that these
peaks are not at the expected position.

\begin{figure} \includegraphics[width=.47\textwidth]{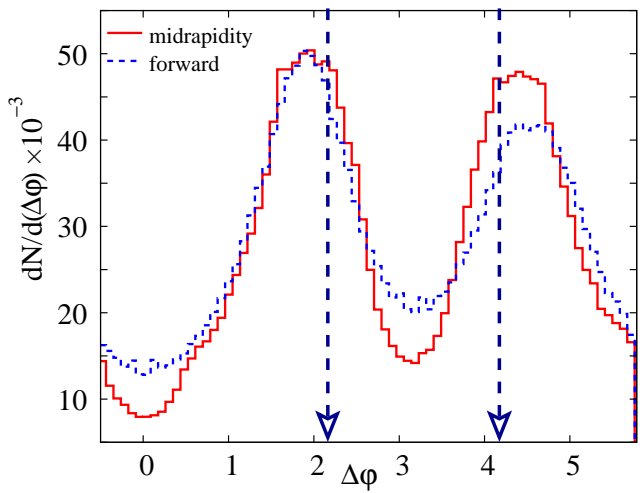}
\caption{(Color Online) Two-particle correlations (backward hemisphere) for
central LHC-collisions for forward- and midrapidity-jets. Both
lines show the average over 250 jets from arbitrary positions. The arrows
show the ``perfect mach positions'' $\Delta \varphi = \pi \pm \cos^{-1}
c_\textrm{S}$.}\label{fig:lhc_rt} \end{figure}

A trigger on the direction of the jet (see Fig.~\ref{fig:lhc_rt}) gives the same
picture. Both the midrapidity jets and the forward jets result in
correlation functions with peaks at the same positions as the peaks from the
untriggered data. Please note that since all plots are made for central
collisions, in which the system is azimuthally symmetric, an additional
trigger on in-/out-of-plane-jets is meaningless.

While triggering on the rapidity of a jet is experimentally a very easy
task, it is impossible to model-independently measure the origin of a jet
inside the reaction zone. But since all data with averaged origins show the
same behaviour, it is interesting to look for cases where the origin is
fixed.

\begin{figure} \includegraphics[width=.47\textwidth]{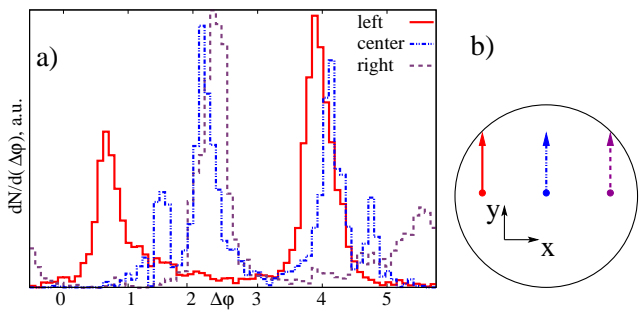}
\caption{(Color Online) (a): Two-particle correlations (backward hemisphere)
for central LHC-collisions for single midrapidity-jets started from
different positions in the collision zone. (b): The
starting positions for the jets shown in (a).}\label{fig:lhc_lmr}
\end{figure}

In Fig.~\ref{fig:lhc_lmr}~(a) the correlation functions for three different
jet origins (which are shown in Fig.~\ref{fig:lhc_lmr}~(b)) are shown. All
jets go in midrapidity and into the same azimuthal direction. The latter
should not make any difference due to the symmetry present in central
collisions. 

The figure shows how much the correlation functions differ. While the
correlation function from the centrally started jet is both symmetric around
$\pi$ and has the peaks at the expected positions $\Delta \varphi \approx
\pi \pm 0.96$, the more peripherally started jets have one of the peaks much
smaller than the other and far away from the expected position. When
averaging over lots of jets coming from peripheral positions, it is obvious
why the correlation function looks like the ones shown in
Figs~\ref{fig:rhic_00},~\ref{fig:lhc_mb}~and~\ref{fig:lhc_rt}. 

One important result is that no correlation function shows the
``wide-peak''-structure as observed at RHIC.

\section{Conclusions}

MACE shows that if sound waves are created by a jet propagating through
Quark-Gluon-Plasma, two- and three-particle correlation functions will show
the expected double- and fourfold-peaked structures. The cone is not washed
out by collective flow.

The peaks will, though, not be at the expected ``perfect mach positions''
$\Delta \varphi_{1,\,2} = \pi \pm \cos^{-1} c_{\rm S}$, but shifted to
positions expected for lower sound velocities. This shift of the maxima is
strong enough ($\delta \Delta \varphi_{\rm max} \approx 0.2$) to feign the
presence of Hadron gas in favour of the (actually present)
Quark-Gluon-Plasma.

The presence of a double-peaked structure (for two-particle correlations)
means that no wide-peak structure arises in MACE. That means that it is not
created by an enormous shift of the conical angles by flow. Non-linear
effects in the cone which might occur (spatially) near the jet may be a
better explanation. If the created shape is not a true cone, but a cone with
a round top, particle emission will be enhanced in backward region with
respect to the results from MACE. Also, a slowing down of the jet will
change the correlation signal to enhance the backward region.

Event-by-event analysis might give better insight into the real speed of
sound. Correlation functions from mid-rapidity jets that are symmetric around
$\pi$ show their peaks at the expected positions. But single-event triggers
may be useless due to event-by-event fluctuations.

\acknowledgements

This work has been supported by the German National Academic Foundation.
B.~B\"auchle wishes to thank Hannah Petersen, Barbara Betz, Leonid Satarov,
Marcus Bleicher and Igor Mishustin for fruitful discussions.

\bibliography{biblio} 

\end{document}